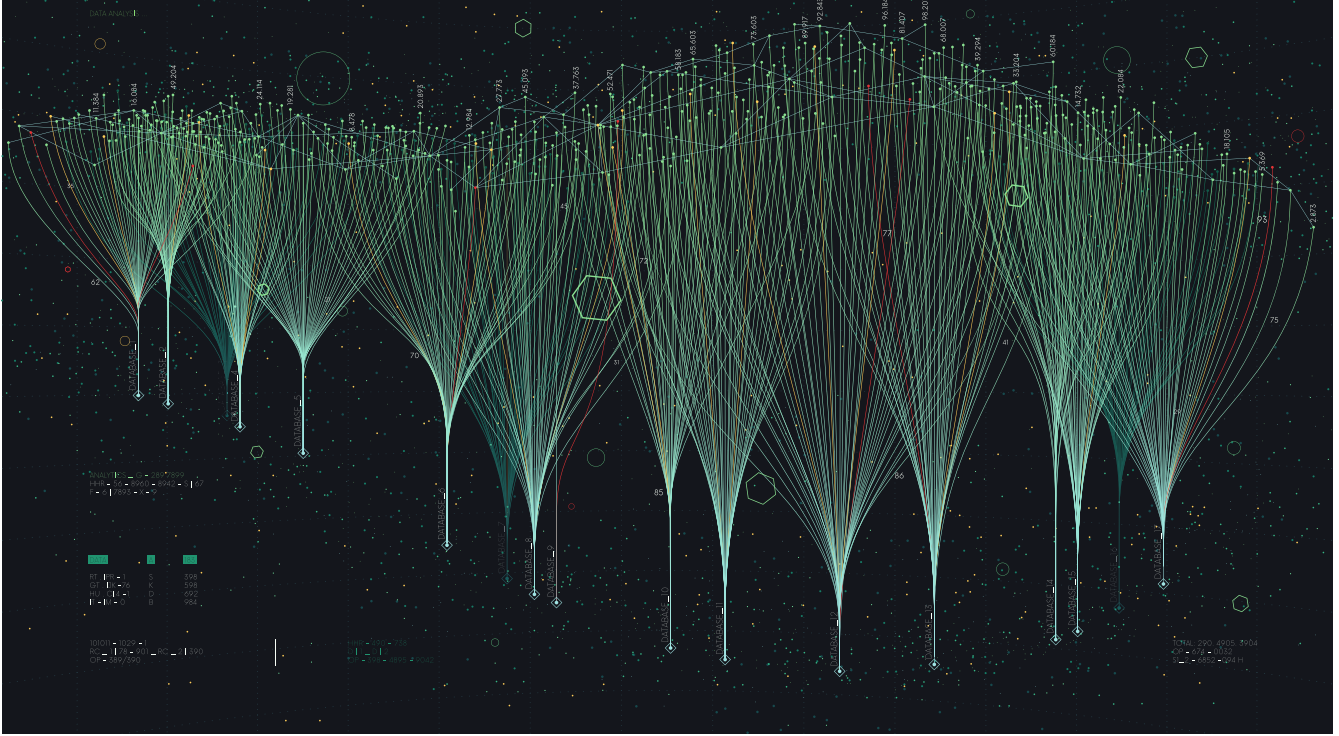

Bild: Shutterstock

**Ein Wegweiser durch den Dschungel analytischer Datenarchitekturen**

# Von Data Warehouse bis Data Mesh

Ein Beitrag von
Torsten Priebe,
Sebastian Neumaier
und Stefan Markus

Data Warehouse, Data Lake, Date Lakehouse, Data Mesh … in der Szene kursieren derzeit viele neue Namen für analytische Datenarchitekturen. Doch sind die diversen Ansätze wirklich so unterschiedlich? Dieser Beitrag versucht einen strukturierten Vergleich der verschiedenen Architekturparadigmen, methodisch basierend auf DAMA-DMBOK und ArchiMate. Es werden Unterschiede, Gemeinsamkeiten und Abhängigkeiten sowie überlappende Architekturbausteine herausgearbeitet und illustriert. Daraus entsteht eine erste Orientierungshilfe für die Wahl der richtigen analytischen Datenarchitektur für den jeweiligen Anwendungsfall.

**Abb. 1:** Betrachtete Architekturparadigmen und ihre Entstehung im Zeitablauf

Data Science und maschinelles Lernen sind derzeit in aller Munde. Da sinnvolle Analysen nur auf einer soliden Datenbasis möglich sind, werden Architekturen wie Data Fabric oder Data Mesh vorgeschla-gen. Aber sind diese „modernen" Ansätze wirklich so anders als „traditionelle" wie Data Warehouse? In diesem Beitrag geben wir einen systematischen Überblick über die verschiedenen analytischen Datenarchitekturen, die in den vergangenen Jahrzehnten vorgeschlagen und etabliert wurden. Um die Gemeinsamkeiten und Unterschiede zu verstehen, schlagen wir eine Struktur in ArchiMate-Notation [Ope 19] vor, die auf den wichtigsten Elementen des „Data Management Wheel" im DAMA-DMBOK [DAM 17] beruht.

Wir behandeln Gartners Logical-Data-Warehouse- und Data-Fabric-Konzepte [EdjB 12, Zai 19] sowie den Data-Mesh-Vorschlag von Dehghani [Deh 19] im Detail. Vergleichend betrachten wir auch das klassische Data Warehouse, Linstedts Data Vault, den Data Lake, Lambda- und Kappa-Architekturen sowie das von Databricks geprägte Data Lakehouse [Inm 21]. All diese Architekturvorschläge finden sich in Abbildung 1 auf einer (ungefähren) Zeitleiste nach ihrer Entstehung beziehungsweise Veröffentlichung aufgetragen. Auf die verschiedenen ArchiMate-Symbole und die eingezeichneten Abhängigkeiten gehen wir im Folgenden noch näher ein.

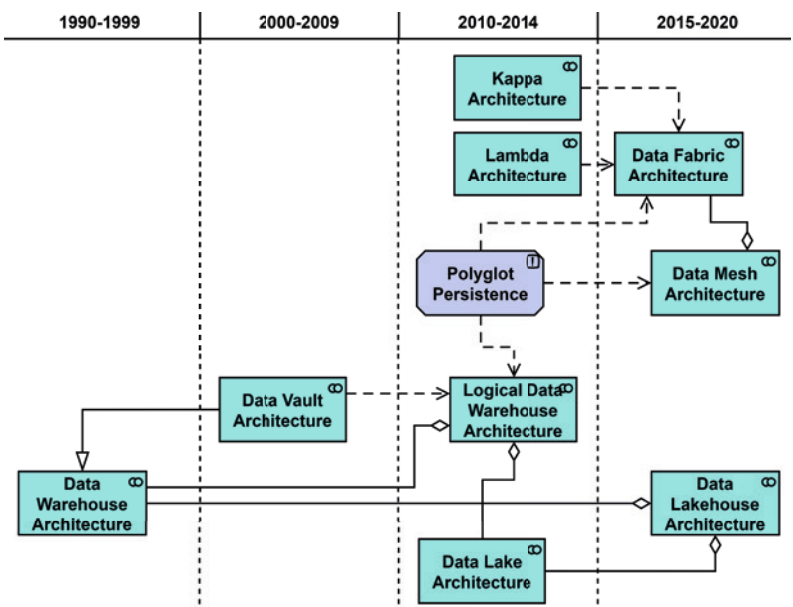

| 1990-1999 | 2000-2009 | 2010-2014 | 2015-2020 |
|---|---|---|---|

Bei dieser Vielzahl von Datenarchitekturvorschlägen ist es schwierig, den Überblick zu behalten. Aus der Softwareentwicklung kennt man das Konzept der Muster und Mustersysteme (Patterns). Wir versuchen hier, einen solchen Ansatz für Datenarchitekturen zu etablieren, um einen Wegweiser für die Wahl des richtigen Architekturparadigmas für die richtige Situation zu bieten. Es würde den Umfang dieses Beitrags jedoch sprengen, alle genannten Architekturvarianten im Detail darzustellen. Daher sei bei einigen auf die unterstützende (noch im Aufbau befindliche) Website unter https://patterns.dataintelligence.blog verwiesen.

## Architekturmuster auf Basis von DAMA-DMBOK und ArchiMate

Wie bereits erwähnt, stützen wir unseren strukturellen Rahmen auf DAMA-DMBOK und ArchiMate. In Abbildung 2 repräsentieren wir die DAMA-Elemente „Data Integration & Interoperability", „Data Storage & Operations", „Data Quality", „Data Security" und „Metadata" in ArchiMate als *Funktionen* der *Application Architecture* einer Datenarchitektur, die wiederum als *Collaboration* dargestellt wird. Wir haben uns entschieden, das DAMA-Element „Data Warehousing & Business Intelligence" nicht aufzunehmen, da wir ein Data Warehouse als eine konkrete Komponente in einer entsprechenden Architektur sehen. Außerdem ergänzen wir den Rahmen mit „Business Intelligence & Data Science", denen ein eigenes Kapitel im DAMA-DMBOK gewidmet ist, auch wenn es kein entsprechendes Segment im Rad gibt [DAM 17]. „Data Governance" und „Data Modeling & Design" werden als ArchiMate *Capabilities* dargestellt, da sie keine Funktionen der Architektur, sondern eher unterstützende Fähigkeiten der Organisation sind. Wir ergänzen außerdem „Data Sources" (als *Datenobjekt*) und Datenkonsumenten (als *Actors*) sowie den Datenfluss.

## Data Warehouse, Data Vault und Data Lake

Data Warehouse (DWH), Data Vault und den mittlerweile ebenso „klassischen" Data Lake betrachten wir hier aus Platzgründen nicht im De-

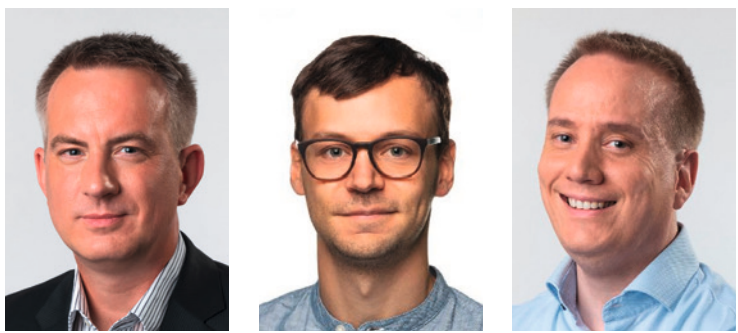

**DR. TORSTEN PRIEBE** ist Dozent für Big Data Analytics an der Fachhochschule St. Pölten und leitet dort die Forschungsgruppe Data Intelligence. Zuvor verantwortete er den Bereich Business Intelligence bei Capgemini in Wien, arbeitete als Solution Architect bei Teradata und übernahm später die Position des CTO bei Simplity.
**E-Mail: torsten.priebe@fhstp.ac.at**

**DR. SEBASTIAN NEUMAIER** ist Researcher in der Forschungsgruppe Data Intelligence an der Fachhochschule St. Pölten. In seiner Doktorarbeit beschäftigte er sich mit der Integration und Anreicherung von Open-Data-Quellen mit Hilfe von Knowledge-Graph-Technologien. Seine aktuelle Forschung konzentriert sich auf verschiedene Aspekte des semantischen Datenmanagements.
**E-Mail: sebastian.neumaier@fhstp.ac.at**

**STEFAN MARKUS** ist Head of Professional Services bei Simplity, einem auf Datenmanagement und Analytics spezialisierten Beratungshaus mit Standorten in Wien, Prag und Brünn. Dort ist er als Solution Architect in spannenden Projekten involviert, vornehmlich im Finanzdienstleistungssektor und E-Commerce. Zuvor war er als Berater bei Teradata in Wien tätig.
**E-Mail: stefan.markus@simplity.ai**

tail. Entsprechende Architekturmuster – aufbauend auf der im vorigen Abschnitt dargestellten Struktur – werden sich auf der genannten Website finden. Dort wird auch eine Unterscheidung

**Abb. 2:** Struktureller Rahmen auf der Grundlage von DAMA-DMBOK und ArchiMate

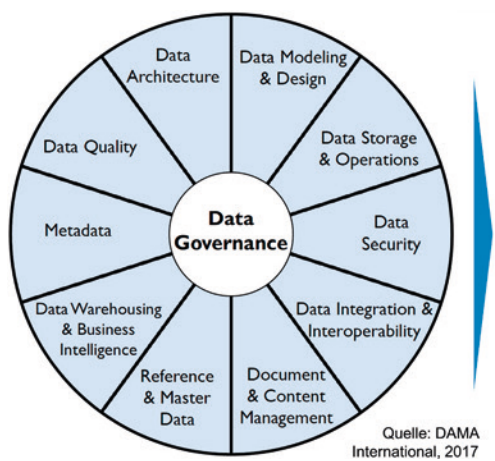

Quelle: DAMA International, 2017

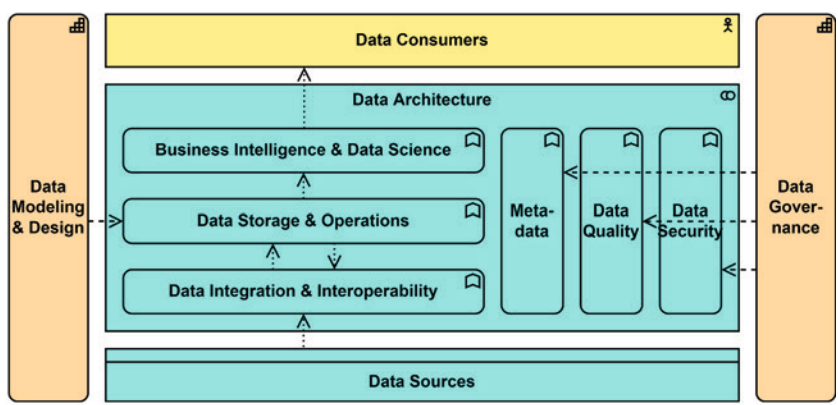

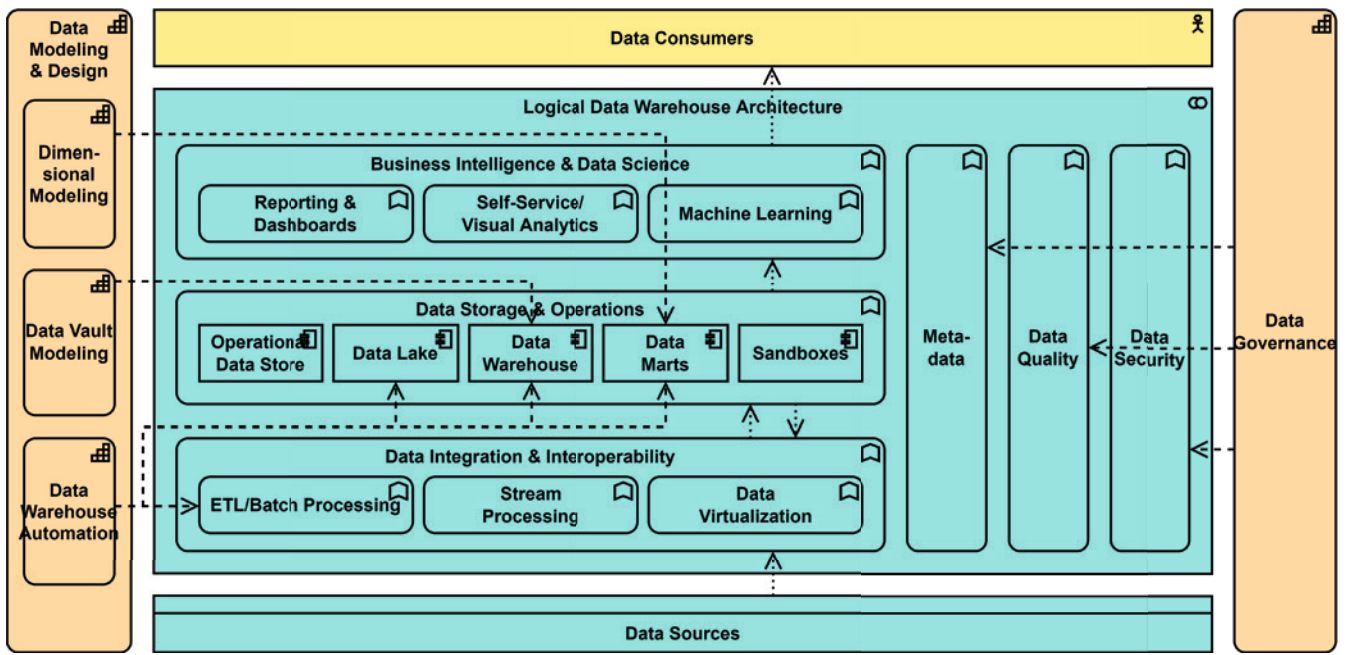

zwischen DWHs nach Kimball und Inmon darge-
stellt.

Auch wenn sich Varianten wie (Near) Realtime
DWH und dergleichen mit Stream-Verarbeitung
herausgebildet haben, erfolgt die Datenintegrati-
on beim DWH in der Regel mit batchbasierter ETL-
(beziehungsweise ELT-) Verarbeitung. Das DWH
basiert speichertechnisch auf relationaler Daten-
banktechnologie. Wie in Abbildung 1 dargestellt,
sehen wir die Data-Vault-Architektur als Spezia-
lisierung der DWH-Architektur. Im Gegensatz da-
zu verwendet ein Data Lake üblicherweise einen
Datei-/Objektspeicher (Hadoop beziehungsweise
AWS S3 oder Azure Blob Store in der Cloud). Bei
der Datenintegration kommt die Stream-Verarbei-
tung hinzu.

### Logical Data Warehouse

Darauf aufbauend wollen wir in diesem Beitrag
das Konzept des Logical Data Warehouse als
erstes Architekturparadigma näher betrachten.
Es wurde von Gartner im Jahr 2012 vorgestellt
[EdB 12] und gibt Empfehlungen, wie Unterneh-
men ein bedarfsgerechtes Datenmanagement
für analytische Anwendungen aufbauen können.
Laut den Autoren sind architektonische Ansätze
wie Data Warehouse, Data Lake und Datenvirtu-
alisierung nicht als konkurrierende Lösungen zu
verstehen, sondern als komplementäre Kompo-
nenten einer übergreifenden Architektur. Gart-
ner erwähnt auch explizit Sandboxes und Stream
Processing als Komponenten einer Logical-DWH-
Architektur. Dies ist in Abbildung 3 entsprechend
dargestellt.

In Bezug auf Datenmodelle nennt Gartner
die dimensionale Modellierung und die Data-
Vault-Modellierung. Es gibt einen Fokus auf Data
Warehouse Automation, weshalb wir diese auch
als *Capability* im Bereich „Data Modeling & De-
sign" mit aufnehmen.

### Polyglot Persistence, Lambda- und Kappa-Architektur

Die Kombination mehrerer Repositories, das heißt
verschiedener Datenspeichertechnologien, wurde
bereits unter der Bezeichnung „Polyglot Persis-
tence" propagiert. Dieser Begriff wurde offenbar
erstmals 2008 von Scott Leberknight verwendet
und 2011 von Martin Fowler aufgegriffen. Er be-
deutet im Grunde, dass jedes Unternehmen mitt-
lerer Größe Technologien wie verteilte Datei- oder
Objektspeichersysteme, relationale und Graph-Da-
tenbanken je nach Bedarf sogar innerhalb einer
einzigen Anwendung kombinieren sollte. Polyglot
Persistence ist dabei eher ein *Prinzip* (und entspre-
chend in ArchiMate-Notation dargestellt), das die
Verwendung eines Polyglot Data Store, sprich einer
Datenspeicherlösung, welche verschiedene Spei-
chertechnologien kombiniert, als *Architekturkom-
ponente* befürwortet. Dies ist entsprechend in Ab-
bildung 4 verzeichnet.

Abbildung 4 zeigt außerdem die Lambda- und
Kappa-Architektur, auf die wir hier wiederum nicht
im Detail eingehen können. Es sei nur so viel er-
wähnt, dass es sich bei den beiden nicht um voll-
ständige Datenarchitekturen handelt, sondern
um einen Fokus auf die Datenintegrationsfunkti-
on. Lambda kombiniert die Batchverarbeitung in
einem sogenannten Batch Layer mit der Stream-
Verarbeitung in einem sogenannten Speed Layer,
während Kappa ausschließlich auf Streaming
setzt (gegebenenfalls durch Log-Verarbeitung oder
Change Data Capture). Beide Architekturen wur-
den 2014/15 entwickelt und veröffentlicht.

### Data Lakehouse

Auch wenn der Begriff „Data Lakehouse" anschei-
nend bereits früher erste Verwendungen gefunden
hat, erfolgte die deutlichste Prägung in einem Blog
von Databricks aus 2020 und darauf aufbauend

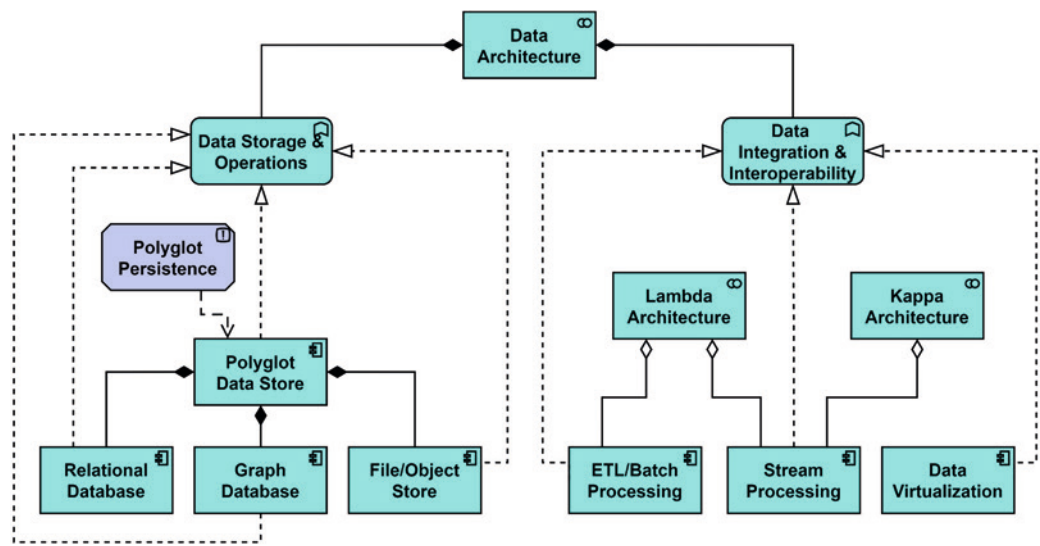

(ebenfalls von Databricks gesponsert) in Inmon et al. [Inm21]. Ein Data Lakehouse ist danach „ein neues, offenes Paradigma, das die besten Elemente von Data Lakes und Data Warehouses kombiniert". Insofern ist die Grundidee mit dem zuvor beschriebenen Logical Data Warehouse nach Gartner vergleichbar. Im Gegensatz dazu setzt das Data Lakehouse (zumindest bei Databricks) jedoch nicht auf die Kombination mehrerer Speichertechnologien, sondern auf einen um Transaktionskonsistenz erweiterten Datei-/Objektspeicher. Da die Grenzen fließender sind als bei logischen DWH, wird nicht zwischen Data Lake und DWH als Architekturkomponenten, sondern zwischen Datensätzen mit Rohdaten und aufbereiteten Daten („Curated Data") unterschieden. Ein entsprechendes Architekturmuster zum Data Lakehouse lässt sich ebenfalls auf der oben erwähnten Website finden.

## Data Fabric und Data Mesh

Die Kombination verschiedener Datenspeicherungs- und -integrationstechniken, jedoch ohne Beschränkung auf konkrete Architektur-Archetypen wie Data Lake oder Data Warehouse, führte zu dem Begriff „Data Fabric", der ursprünglich im Jahr 2015 von George Kurian von NetApp geprägt und dann von Gartner [Zai19] wieder aufgegriffen wurde. Laut den Autoren ist Data Fabric „ein Designkonzept zur Erreichung wiederverwendbarer und erweiterter Datenintegrationsdienste, Datenpipelines und Semantik mit dem Ziel einer flexiblen und integrierten Datenbereitstellung". Data Fabric kann als Nachfolger und Verallgemeinerung des Logical DWH gesehen werden. Sie baut auf der Idee der Polyglot Persistence beziehungsweise einem Polyglot Data Store auf, der Speicheransätze wie relationale Datenbanken, Graph-Datenbanken und/oder Datei-/Objektspeicher kombiniert.

Ein Hauptaugenmerk des Data-Fabric-Konzepts liegt auf Metadaten, die laut Gartner aus einem (erweiterten) Datenkatalog und einem Knowledge-Graph bestehen, der semantisch verknüpfte Metadaten enthält. Der Einsatz von Künstlicher Intelligenz beziehungsweise maschinellem Lernen zur teilweisen Automatisierung der Metadatenerstellung wird von Gartner mit dem Begriff „aktive Metadaten" bezeichnet. Data Fabric basiert außerdem auf der Grundidee, Daten als Dienste anzubieten. Gesponsert von Denodo, unterscheidet van der Lans [Lan21] transaktionale Dienste, die mit klassischen datensatzbasierten APIs umgesetzt werden, und analytische Dienste, die flexible Anfragen per SQL unterstützen, für die sich eben Datenvirtualisierung anbietet. Im Gegensatz zum Logical DWH werden Reporting- und Analyse-Werkzeuge nicht als Kern des Data-Fabric-Konzepts gesehen, sondern fallen in den Verantwortungsbereich der Datenkonsumenten.

Data Fabric führt damit im Grunde bereits die Grundidee von „Data Products" ein, auch wenn diese Terminologie nicht explizit verwendet wird. Daraus resultierte schließlich die Idee eines Data Mesh nach Dehghani [Deh19]. Dehghani argumentiert, dass die bestehenden zentralisierten und monolithischen Datenverwaltungsplattformen, die keine klaren Domänengrenzen und Eigentum an Domänendaten kennen, bei großen Unternehmen mit einer vielfältigen Zahl von Datenquellen und -konsumenten versagen. In einem Data Mesh müssen die Domänen ihre Datensätze als Domänendaten selbst hosten und als Datenprodukte bereitstellen. Während die einzelnen Domänenteams die Technologie zur Speicherung, Verarbeitung und Bereitstellung ihrer Datenprodukte eigenständig kontrollieren, sorgt eine gemeinsame Plattform für die einheitliche Interaktion mit den Datenprodukten.

Abbildung 5 zeigt dieses Paradigma durch Hinzufügen von Business Domains (dargestellt als ArchiMate *Business Collaborations*) und ihre Domänen-Datenprodukte mit ihren *Schnittstellen*. Wie im Diagramm zu sehen ist, gehören die Data Pipelines auch in den Besitz der Business-Domänen, das heißt, jede Domäne ist für ihre eigenen Datentransformationen verantwortlich. Eine Domäne kann Datenprodukte aus einer anderen Domäne konsumieren. Wie bei Data Fabric liegt ein Schwerpunkt auf Metadaten mit einem Datenkatalog, der ein domänenübergreifendes Inventar der verfügbaren Daten-produkte darstellt. Ebenso wie bei Data Fabric gehören Berichts- und Analyse-Tools nicht zum Fokus (daher liegt „Business Intelligence & Data Science" außerhalb der Data-Mesh-Architecture-Box). Jedoch bezieht das Data-Mesh-Konzept im Gegensatz zu den anderen vorgestellten Datenarchitektur-Paradigmen die Datenquellen in die Betrachtung mit ein. Operative Daten werden über operative Datenprodukte (beziehungsweise deren Schnittstellen) bedient, genauso wie analytische Datenprodukte.

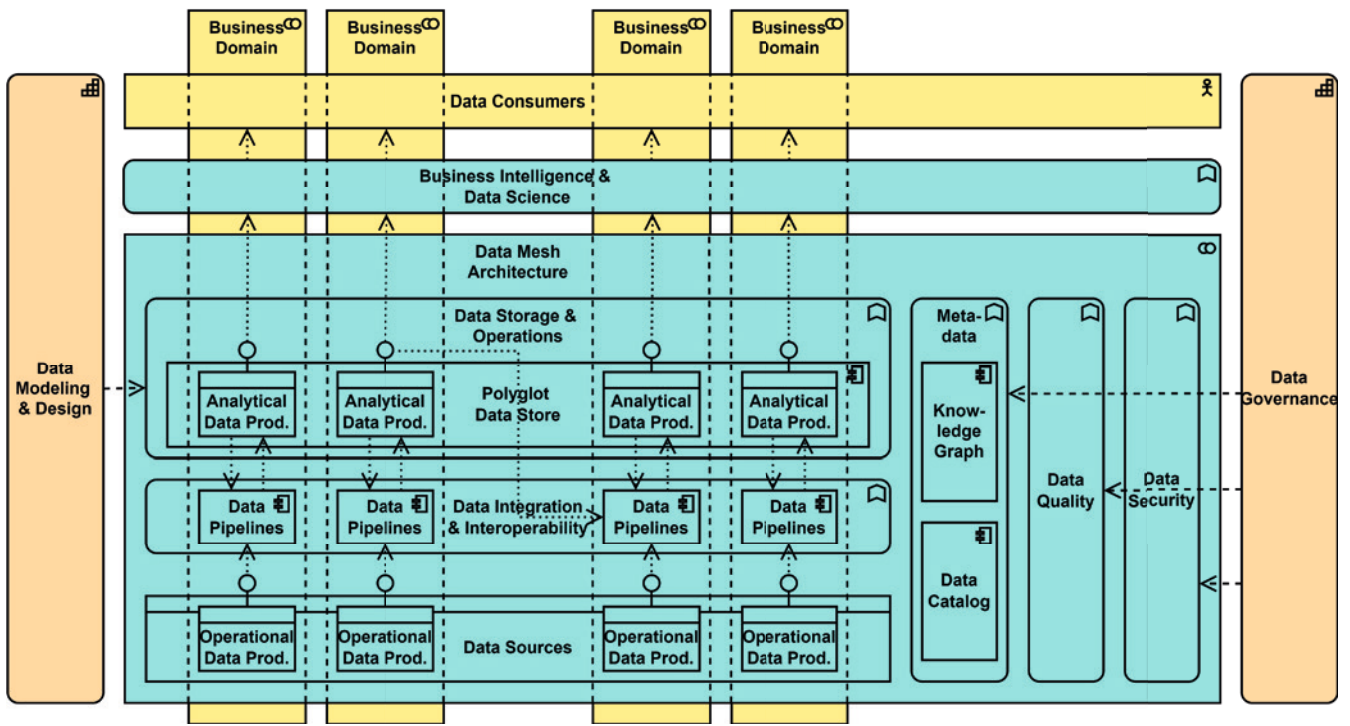

## Vergleich und Fazit

In diesem Beitrag haben wir eine erste systematische Darstellung wichtiger analytischer Datenarchitekturen in einem gemeinsamen strukturellen Rahmen basierend auf einer gemeinsamen semiformalen Notation vorgenommen. Wir haben dafür das DAMA-DMBOK und ArchiMate verwendet. Es wurden insbesondere Logical-Data-Warehouse- und Data-Mesh-Architekturen detaillierter behandelt.

In Anlehnung an Inmon et al. [Inm21] wagen wir hier einen ersten Vergleich nach den Dimensionen „Datenformat", „Datentypen", „Datenzugriff", „Zuverlässigkeit", „Governance und Sicherheit", „Performance", „Skalierbarkeit" und „unterstützte Anwendungsfälle". Wir ergänzen „Datenaktualität", da sich die Architekturvarianten auch in ihrer Unterstützung von Streaming und Datenvirtualisierung signifikant unterscheiden. Die in der Quelle dargestellte Originaltabelle, die sich auf Data Warehouse, Data Lake und Data Lakehouse beschränkt, ist naturgemäß etwas in Richtung Data Lakehouse gefärbt. Wir haben in Tabelle 1 versucht, das etwas zu objektivieren.

Auf der oben genannten Website werden wir weitere Architekturen und Paradigmen wie Lambda- und Kappa-Architekturen im Detail abdecken und als Architekturmuster darstellen – bis hin zu einem Mustersystem ähnlich den Software-Entwurfsmustern der „Gang of Four" [Gam95]. Dort werden sich auch detaillierte Mustervorlagen mit Kontext-, Problem- und Lösungsabschnitten finden, die dann hoffentlich eine noch bessere Orientierungshilfe für die Auswahl der richtigen Architekturparadigmen bieten.

## Literatur

| | Data Warehouse | Data Vault | Data Lake | Logisches Data Warehouse | Data Fabric | Data Mesh | Data Lakehouse |
|---|---|---|---|---|---|---|---|
| **Datenformat** | Relationale Datenbank, physische Speicherung proprietär | Relationale Datenbank, physische Speicherung proprietär | Datei-/Objekt-speicher auf Basis offener Dateiformate | Unterschiedliche Datenformate mit Polyglot Persistence | Unterschiedliche Datenformate mit Polyglot Persistence | Unterschiedliche Datenformate mit Polyglot Persistence | Datei-/Objekt-speicher auf Basis offener Dateiformate |
| **Datentypen** | Strukturierte Daten, begrenzte Unterstützung für semi-strukturierte Daten | Strukturierte Daten, begrenzte Unterstützung für semi-strukturierte Daten | Strukturierte Daten, semi-strukturierte Daten, textuelle Daten, unstrukturierte (Roh-)Daten | Strukturierte Daten im DWH, semi-strukturierte, textuelle und unstrukturierte Daten im Data Lake | Strukturierte Daten, semi-strukturierte Daten, textuelle Daten, unstrukturierte (Roh-)Daten | Strukturierte Daten, semi-strukturierte Daten, textuelle Daten, unstrukturierte (Roh-)Daten | Strukturierte Daten, semi-strukturierte Daten, textuelle Daten, unstrukturierte (Roh-)Daten |
| **Daten-aktualität** | Üblicherweise täglich, Limitierung durch Batch-/ETL-Zyklen | Üblicherweise täglich, Limitierung durch Batch-/ETL-Zyklen | Nahezu Echtzeit durch Streaming möglich | Batch-/ETL-Zyklen im DWH, Streaming im Data Lake möglich | Bis zu Echtzeit möglich durch Datenvirtuali-sierung (je nach Dienst) | Bis zu Echtzeit möglich durch Datenvirtuali-sierung (je nach Datenprodukt) | Kombination von Batch-/ETL-Zyklen und Streaming möglich |
| **Datenzugriff** | SQL | SQL | Offene APIs für Datei-/Objektzugriff, begrenzter SQL-Zugriff | Offene APIs für Datei-/Objektzugriff im Data Lake, SQL im DWH und begrenzt übergreifend mit Daten-virtualisierung | Offene APIs für transaktionale Dienste, SQL (mit Daten-virtualisierung) für analytische Dienste | Offene APIs für operative Datenprodukte, SQL (mit Daten-virtualisierung) für analytische Datenprodukte | Offene APIs für Datei-/Objekt-zugriff, SQL |
| **Zuver-lässigkeit** | Hohe Qualität und Zuverlässig-keit, ACID-Transaktionen | Mittlere Qualität im Raw Vault, hohe Qualität im Business Vault, ACID-Transaktionen | Geringe Qualität und Zuverlässigkeit | Geringe Qualität und Zuver-lässigkeit im Data Lake, hohe Qualität, ACID-Transaktionen im DWH | Abhängig vom Dienst | Abhängig vom Datenprodukt | Geringe Qualität und Zuverlässigkeit bei Rohdaten, hohe Qualität, ACID-Trans-aktionen bei aufbereiteten Daten |
| **Data Governance und Sicherheit** | Feingranulare Sicherheit und Governance (Zeilen-/Spalten-ebene) | Feingranulare Sicherheit und Governance (Zeilen-/Spalten-ebene) | Schwache Sicherheit und Governance, da Dateiebene | Feingranular im DWH, schwache Sicherheit und Governance im Data Lake | Abhängig vom Dienst | Abhängig vom Datenprodukt | Feingranular bei SQL-Zugriff, schwache Sicherheit und Governance bei Datei-/Objekt-zugriff |
| **Performance** | Hoch, da spezifisch optimierbar | Hoch, da spezifisch optimierbar | Eher niedrig, da datei-/objektbasiert, abhängig von der Daten-verwendung (MapReduce, Spark) | Hoch im DWH, eher niedrig im Data Lake (abhängig von der Daten-verwendung) | Abhängig vom Dienst | Abhängig vom Datenprodukt | Mittel, da bis-lang begrenzte Optimierungs-möglichkeiten |
| **Skalierbarkeit** | Skalierung wird exponentiell teurer | Skalierung wird teurer, aber leichter zu handhaben als im klassischen DWH | Hoch skalierbar für große Datenmengen bei niedrigen Kosten | Skalierung teuer im DWH, günstig im Data Lake | Abhängig vom Dienst | Abhängig vom Datenprodukt | Hoch skalierbar für große Datenmengen bei niedrigen Kosten |
| **Unterstützte Anwendungs-fälle** | Klassisches BI, Berichte, Dash-boards, SQL | Klassisches BI, Berichte, Dash-boards, SQL | Data Science, insbesondere maschinelles Lernen | Vielfältig, von klassischem BI über Self-Service-BI bis zu maschinellem Lernen | Je nach Datenprodukt, klassisches BI, Self-Service-BI, maschinelles Lernen, auch operative Anwendungen | Je nach Datenprodukt, klassisches BI, Self-Service-BI, maschinelles Lernen, auch operative Anwendungen | Vielfältig, von klassischem BI über Self-Service-BI bis zu maschinellem Lernen |

**Tab. 1:** Betrachtete Architekturparadigmen im Vergleich